# Big5PersonalityEssays: Introducing a Novel Sythetic Generated Dataset Consisting of Short State-of-Consciousness Essays Annotated Based on the Five Factor Model of Personality

Iustin Floroiu[1,2]

[1]University POLITEHNICA of Bucharest

[2]The National Institute for Research and Development in Informatics – ICI Bucharest

Bucharest, Romania

iustin.floroiu@yahoo.com

**Abstract**: Given the high advances of large language models (LLM) it is of vital importance to study their behaviors and apply their utility in all kinds of scientific fields. Psychology has been, in recent years, poorly approached using novel computational tools. One of the reasons is the high complexity of the data required for a proper analysis. Moreover, psychology, with a focus on psychometry, has few datasets available for analysis and artificial intelligence usage. Because of these facts, this study introduces a synthethic database of short essays labeled based on the five factor model (FFM) of personality traits.

**Keywords**: Dataset, state-of-consciousness essays, Large Language Models, Psychometry, Five Factor Model of Personality

## 1.Introduction

Psychology, with a focus on psychometry, heavily relies on statistical models of analysis to create a cohesive understanding of personality and preferences of individuals. One of the fields that showed a proper evolution over time is the psychology of personality. Statistics-wise, personality can be modeled using the Five Factor Model (FFM), this model being the most scientifically validated personality model to date. It consists of five personality traits, each divided into 6 facets, usually. These traits can be memorized using the acronym OCEAN: openness to experience (O), conscienciousness (C) , extraversion (E), agreeableness (A) and neuroticism (N). The traits are not correlated with one another, as evidence suggests. However, these 5 personality traits can be mapped into two metatraits: plasticity and stability. Plasticity is composed of traits openness and extraversion, while stability is consisting of agreeableness, conscienciousness and neuroticism (Costa et al, 1995), (Digman, 1997).

## 2.Describing the Five Factor Model

Each of the five personality traits can be divided into 6 sub-traits, called facets. On this premise, the 5 traits can be described in more depth.

Openness to experience can be described using the facets imagination, artistic interests, emotionality, adventurousness, intellect and liberalism. Individuals high in imagination are creative, original, and have a vivid imagination. They enjoy exploring new ideas and concepts. People with high artistic interests appreciate art, music, and aesthetics. They often engage in creative activities and express themselves through various art forms. High emotionality indicates sensitivity to emotions and a tendency to experience intense feelings. These individuals are often more aware of their feelings and values. Individuals high in adventurousness seek excitement and new experiences. They enjoy taking risks and exploring unfamiliar territory. People with high intellect value intellectual pursuits and enjoy engaging in complex thinking and problem-solving. They tend to be interested in abstract ideas and philosophy. Liberalism reflects openness to unconventional ideas and values. Individuals high in liberalism are tolerant of diversity and open-minded (Allport, 1937).

Conscienciousness can be divided into 6 facets, competence, order, dutifulness, achievement-striving, self-discipline and deliberation. Competent individuals rely in their own qualities and are sure of themselves. People high in orderliness prefer structure and organization in their environment. They are neat, tidy, and methodical in their approach to tasks, besides having a high capacity for planning. Dutiful individuals are responsible, reliable and proeficient at fulfilling their obligations and commitments diligently. Individuals high in achievement-striving are ambitious and motivated to succeed. They set goals for themselves and work hard to accomplish them. Self-disciplined individuals have strong willpower and the ability to control their impulses. They are disciplined in their habits and behaviors. Deliberate individuals are cautious and thoughtful in their decision-making process. They carefully weigh the pros and cons before taking action (Allport, 1937).

Extraversion can be understood more deeply using the facets friendliness, gregariousness, assertiveness, activity level, excitement-seeking and positive emotions. Friendly individuals are sociable, outgoing, and enjoy interacting with others. They are warm and approachable, making them easy to get along with. They tend to be interested in the lives of others. Gregarious individuals thrive in social settings and enjoy being surrounded by others. They are lively and enjoy participating in group activities. They do not mind crowded areas and attention. Assertive individuals are confident and self-assured. They express themselves openly and take charge of situations when necessary. People high in activity level are energetic and lively. They enjoy staying busy and engaging in various activities. Excitement-seeking individuals crave stimulation and excitement. They enjoy taking risks and seeking out thrilling experiences. Positive emotions reflect optimism, enthusiasm, and a cheerful disposition. Individuals high in positive emotions are upbeat and optimistic (Allport, 1937).

Agreeableness can be divided into trust, morality, altruism, cooperation, modesty and sympathy. Trusting individuals believe in the inherent goodness of others and are trusting in their relationships. Moral individuals value honesty, integrity, and ethical principles. They strive to do what is right and treat others with fairness and respect. Altruistic individuals are compassionate and caring toward others. They are generous and willing to help those in need. Cooperative individuals are collaborative in their interactions with others. They are team players and work well in group settings. Modest individuals are humble and down-to-earth. They do not seek attention or praise for their accomplishments and are modest about their abilities. Sympathetic individuals are empathetic and understanding of others' emotions and experiences. They are supportive and compassionate in their relationships(Allport, 1937).

Neuroticism is described based on the following facets: anxiety, anger, depression, self-consciousness, vulnerability and impulsiveness. Anxious individuals are prone to worry, fear, and nervousness. They are sensitive to stress and may experience anxiety-related symptoms. Angry individuals are quick to experience feelings of anger, frustration, and irritability. They may have difficulty controlling their temper and may react impulsively. Depressed individuals experience feelings of sadness, hopelessness, and despair. They may have low self-esteem and struggle with feelings of worthlessness. Self-conscious individuals are overly aware of themselves and how they are perceived by others. They may be sensitive to criticism and self-conscious in social situations. Vulnerable individuals are sensitive to stress and prone to experiencing negative emotions. They may feel overwhelmed by life's challenges and struggle to cope with adversity. Impulsive individuals have difficulty controlling their impulses and may act hastily without considering the consequences of their actions. They may engage in risky behaviors without thinking things through (Allport, 1937).

Eventhough the traits are not correlated to one another, the facets of a trait are believed to present a degree of correlation to one another.

The traits are cuantised for each person, such as an individual can be mapped on a continuous scale of a trait spectrum. Therefore, there are disadvantages and advantages of being lower or higher on a specific scale. However, personality traits tend to be modeled using a normal correlation with regards to population distribution and most people tend to be near the mean rather than farther from it (DeYoung et al, 2002).

Based on this, pros and cons for being higher or lower on a trait with respect to the spectrum of scores has been studied. A brief conclusion based on recent studies can be seen below

## 3.Advantages and Disadvantages of Personality Traits

Starting with Openness to Experience, individuals who score high on this trait are typically imaginative, curious, and open to new experiences, which fosters creativity and adaptability. Such traits are beneficial in roles requiring innovation and problem-solving. However, high openness can also lead to overindulgence in fantasy or impractical ideas, and a lack of focus on routine tasks. On the other hand, low scorers, being more traditional and practical, are often excellent in situations requiring consistency and routine, but they might struggle with adaptability and be less receptive to new ideas, which can be a disadvantage in rapidly changing environments (Goldberg, 1990), (John et al, 1999), (McRae et al, 1992).

Conscientiousness describes a person's level of goal-oriented organization, diligence, and reliability. Highly conscientious individuals tend to be disciplined and excel in following plans and managing details, traits highly valued in many professional settings. The downside can be inflexibility and a propensity to stress when perfection is unattainable or plans are disrupted.

Individuals scoring low on conscientiousness may excel in roles that require flexibility and spontaneity, but may be perceived as unreliable or careless with details (Goldberg, 1990), (John et al, 1999), (McRae et al, 1992).

Extraversion is characterized by high energy, sociability, and a preference for being around others. Extraverts are often seen as charismatic and are generally effective in leadership roles or jobs involving teamwork and networking. However, they may find it challenging to work independently or in quiet environments and can be perceived as overbearing. Conversely, introverts (those with low extraversion scores) often excel in tasks requiring solo work and concentration and may be seen as reflective and observant. Their challenges often lie in networking or active team engagement, which can hinder their visibility in social contexts (Goldberg, 1990), (John et al, 1999), (McRae et al, 1992).

Agreeableness refers to a person's general concern for social harmony and caring for others. Those who score high on this trait are typically cooperative, trustworthy, and compassionate, making them excellent collaborators and mediators. However, they may be prone to being too selfless or unable to engage in conflict when necessary, potentially leading to exploitation. Those low in agreeableness may be more competitive or challenging, qualities that can be advantageous in negotiation scenarios or high-pressure competitive environments. Still, this could also result in difficulties in teamwork and maintaining relationships (Goldberg, 1990), (John et al, 1999), (McRae et al, 1992).

Lastly, Neuroticism measures the tendency toward emotional instability and negative emotions like anxiety and irritability. High scorers on neuroticism may struggle with stress and decision-making under pressure and are more prone to mood swings and burnout. However, their heightened sensitivity to environmental stress can lead to greater caution, potentially mitigating risks effectively. Individuals with low scores in neuroticism generally experience less emotional turmoil and exhibit a stable demeanor, which supports resilience in challenging circumstances. However, their lack of emotional response can sometimes be perceived as detachment or a lack of empathy, particularly in social or collaborative settings (Goldberg, 1990), (John et al, 1999), (McRae et al, 1992).

Each personality trait, therefore, offers distinct advantages and accompanying disadvantages, significantly influencing how individuals interact with the world and achieve success in various aspects of life.

## 4. The Utility of the Dataset

The synthetically generated dataset comprising short state-of-consciousness essays annotated with the Five Factor Model (FFM) of personality encounters significant utility across various domains, notably in psychology, artificial intelligence, and educational sectors. Each of these fields can leverage the dataset to enhance their understanding and application of personality-based data, driving forward their respective methodologies and outcomes.

In the realm of psychology, particularly within personality research, the dataset provides a robust tool for exploring the nuances of personality traits as expressed through written language. By analyzing how different personality traits influence writing style, content, and emotional tone, researchers can gain insights into the cognitive and emotional aspects of personality. This can further the development of personalized therapeutic approaches and enhance the precision in clinical diagnoses related to personality disorders or traits. For example, narcissistic traits are correlated with high extraversion and low agreeableness, while sociopathy traits are usually denoted by lower extraversion (or higher introversion), conscienciousness, agreeableness and neuroticism. Borderline personality traits can be underlined by low extroversion, conscienciousness and agreeableness and high openness and neuroticism (Quinn et al, 2007).

For AI, particularly in natural language processing (NLP) and machine learning (ML), this dataset serves as a training and testing ground for developing algorithms that can effectively interpret and generate text based on underlying personality traits. This capability extends to creating more personalized and context-aware systems, such as chatbots, virtual assistants, or recommendation engines that adapt their responses according to perceived personality traits of users. Such tailored interactions are likely to improve user engagement and satisfaction by offering more empathetic or personality-consistent interactions.

Educationally, the dataset can facilitate the development of tools that adapt teaching methods and materials to the personality traits of students. This can potentially increase learning efficiency and motivation by aligning educational approaches with students' personality-driven learning preferences. Moreover, this insight can guide curriculum developers to create varied content that addresses the diverse personality spectrum within a classroom, ensuring a more inclusive education system that caters to broad learning needs.

In recruitment and human resources, this dataset can enrich personality assessment methods, offering another layer of analysis through how candidates express themselves in writing. This can be particularly useful in positions where communication, creativity, or meticulousness are key job requirements. Thus, HR professionals could better match candidates to roles that suit their personality traits, potentially enhancing job satisfaction and performance.

## 5. Dataset Generation

The dataset was obtained with the aid of the GPT-4-turbo model, using prompt engineering (OpenAI, 2023). Initially, the model was fed with three annotated examples from the Essays dataset. Moreover, the model was instructed, using the prompt, to generate both text and one label for each personality trait, 1 meaning that the trait is above average and 0 below average. The model was instructed to generate a short passage in a role-play manner, behaving as a person with a personality described by the 5 labels, manually prompted by the author, in which it was asked how a day in the life of that person would be. The texts were generated using first person pronouns, confirming that the model behaves as expected using the role-play prompting methodologies. The model generates 3 essays at once, baased on the labels explicitated using the prompt. Every three generations (or after generating 9 essays) the model is fed with a new example from the Essays dataset. After 20 generations (or after 60 essays with labels), another prompt is opened and the process starts again, to make sure all the generated essays are not similar with the ones already generated. The examples from the Essays datased were randomly chosen.

## 6. Results analysis

A number of 400 samples were generated artificially using GPT-4-turbo and prompting methodologies. A comprehensive analysis of the dataset is desired, to prove the utility of the generated prompts.

### 6.1. Descriptive statistics

The first step consisted of a descriptive statistics analysis of the generated data. The results can be observed by using the following table:

|  | Openness | Conscienciousness | Extraversion | Agreeableness | Neuroticism |
|---|---|---|---|---|---|
| count | 399.000000 | 399.000000 | 399.000000 | 399.000000 | 399.000000 |
| mean | 0.536341 | 0.536341 | 0.543860 | 0.553885 | 0.503759 |
| std | 0.499304 | 0.499304 | 0.498698 | 0.497712 | 0.500614 |
| min | 0.000000 | 0.000000 | 0.000000 | 0.000000 | 0.000000 |
| 25% | 0.000000 | 0.000000 | 0.000000 | 0.000000 | 0.000000 |
| 50% | 1.000000 | 1.000000 | 1.000000 | 1.000000 | 1.000000 |
| 75% | 1.000000 | 1.000000 | 1.000000 | 1.000000 | 1.000000 |
| max | 1.000000 | 1.000000 | 1.000000 | 1.000000 | 1.000000 |

*Table 1. The Descriptive Statistics Table of the Dataset*

The dataset indicates a reasonably balanced distribution of labels across the personality traits, with each trait's mean value hovering around the midpoint of 0.5, suggesting a near-equal representation of binary classifications (0 and 1). Specifically, Openness and Conscientiousness both have a mean of 0.536; Extraversion slightly higher at 0.544, Agreeableness at 0.554, and Neuroticism at 0.504. These figures suggest a slight inclination towards a label of 1 (or positive) across most traits, except for Neuroticism which is closer to a perfect balance.

This balance is significant because it indicates a dataset that is not heavily skewed towards any particular personality trait expression. Such balance is essential for training unbiased machine learning models in AI applications, as well as for conducting generalized psychological research where the representation of various personality types affects the validity and reliability of the outcomes.

The descriptive statistics further reveal that all traits adhere to a binary distribution (label values of 0 and 1), with standard deviations close to 0.5. This standard deviation is informative because it indicates a healthy variance within binary data, affirming that the dataset encompasses a diverse set of responses and personality presentations. Such diversity is crucial for depth in psychological analysis and for the development of AI systems capable of recognizing a wide range of human personality traits.

The quartile values illuminate more about the dataset's distribution. With the 25th percentile at 0 and the 75th percentile at 1 for all traits, and the median (50th percentile) consistently at 1, the data aligns well with a median split in terms of personality labeling. This positioning suggests that, despite slight variances, there remains a balanced distribution across observations, which is paramount for conducting studies that require equal representation of personality trait expressions.

A notable observation is the slight skew towards a positive (1) label in Agreeableness, with 55.4% of the data leaning this way. This could suggest a potential bias in the way individuals either perceive themselves or are perceived by others when it comes to agreeable behaviors. In practical settings, such as the workplace or educational institutions, this might imply a preference or a perceived virtue in agreeableness, which could influence teamwork dynamics, leadership styles, and collaborative projects.

6.2. Visual Analysis

To further provide insight into the generated data, the label distribution for all the 5 traits was plotted. Knowing that there are 399 samples for each label, given that each essay contains five binary labels, one for every personality trait, the following graphical representations are believed to explore the dataset in a more concise manner, as follows:

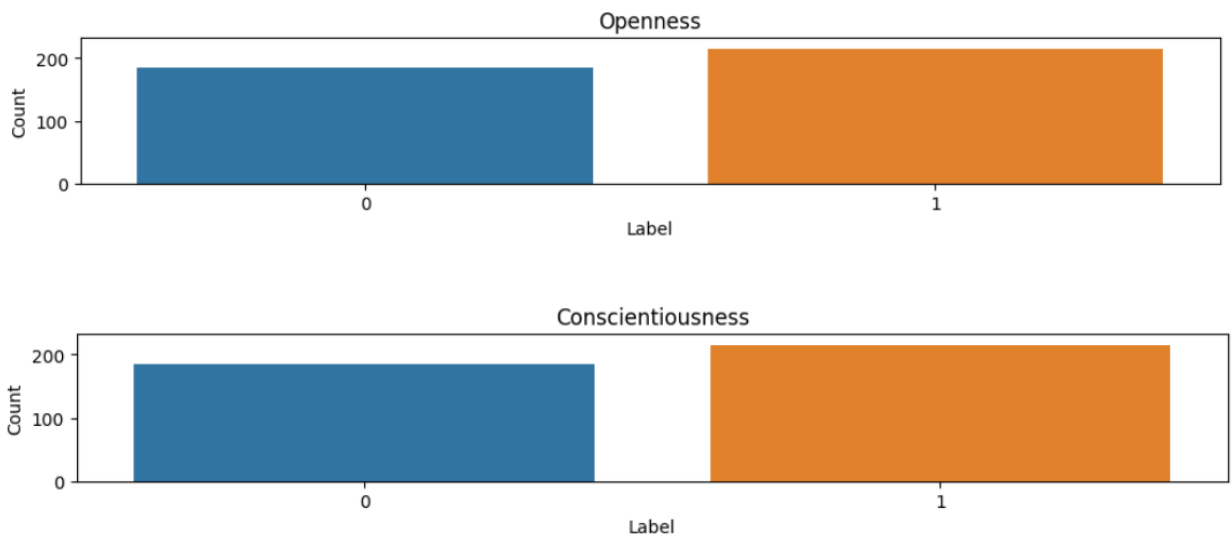

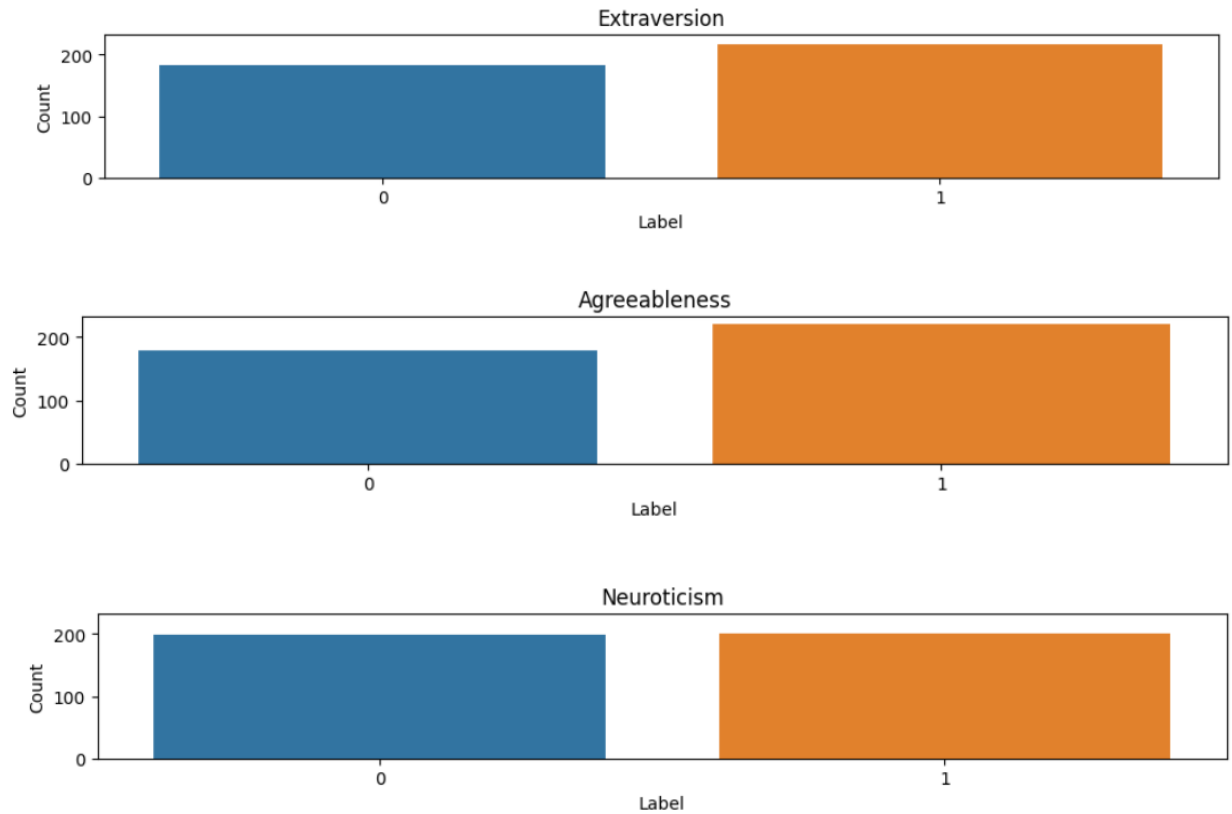

*Figure 1. The Label Distribution of the Generated Data*

## 6.3. Trait Correlation Analysis

For a better usage and more utility, given that, theoretically, the five personality traits are not correlated with one another, it is expected that the data labels are not correlated either. Moreover, lower correlation values (close to 0) are more suitable for training an artificially intelligent model with regards to obtaining accurately validated insight.

The heatmap shown below is believed o enhance the analysis of thedataset from the correlation point of view and susceptibility to artificial intelligent usage and scientific validity:

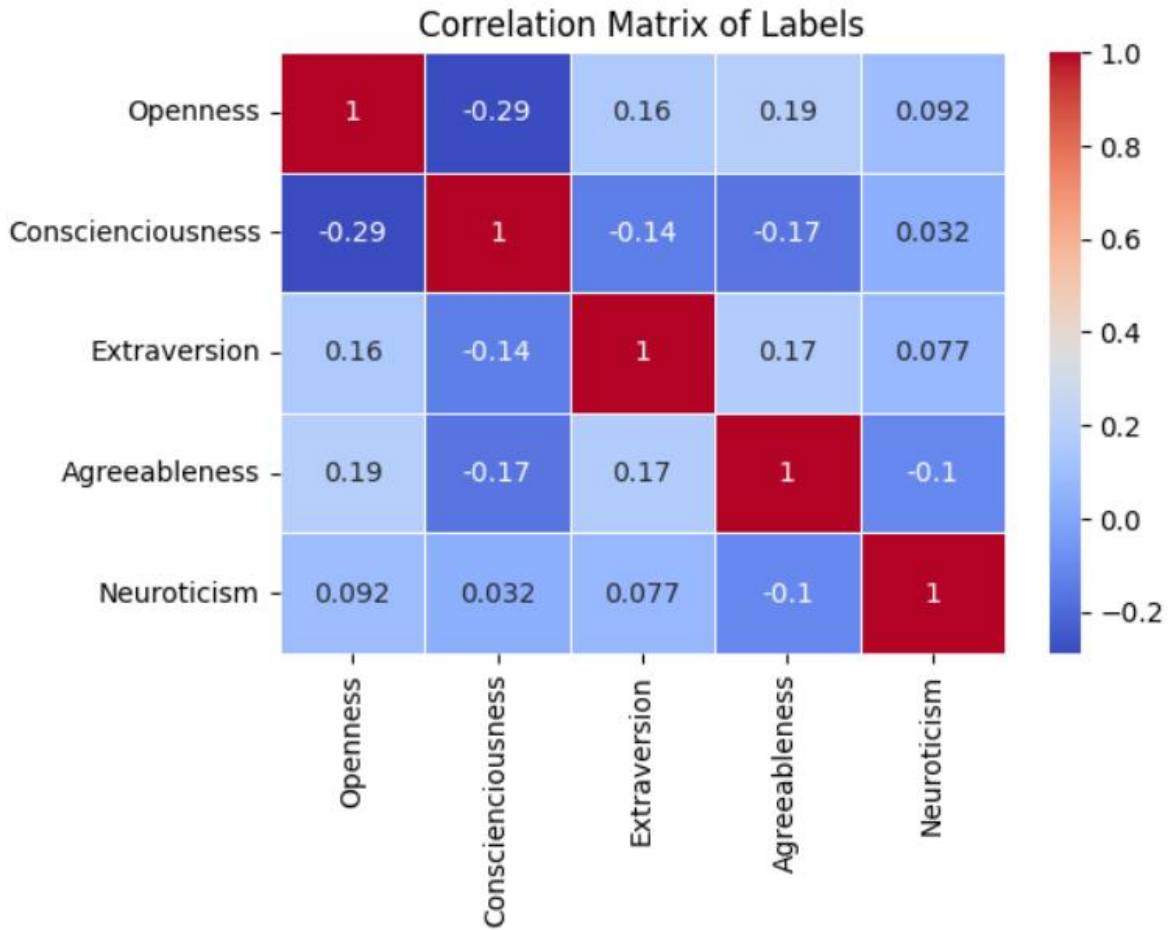

*Figure 2. The Heatmap of the Correlations between labels with regards to the generated text*

As shown in the above figure, it can be seen that the highest correlation value is 0.19, which is low enough to consider the scientific validity of the study for further usage.

## 7.Data Storage

The dataset is currently stored on both GitHub, following the link below.

The following link can be used for accessing the dataset from GitHub:

https://github.com/stnflri/Big5PersonalityEssays

It is strongly advised to cite this paper if the dataset is used with regards to a computational application, study case or for scientific inquiry and analysis.

## 8.Conclusions

In conclusion, the Big5PersonalityEssays dataset represents a significant advancement in the intersection of computational tools and personality psychology. By leveraging the capabilities of large language models such as GPT-4-turbo, this study successfully generated a synthetic dataset of short state-of-consciousness essays annotated with the Five Factor Model of personality traits. This dataset promises considerable utility across various domains, including psychology, artificial intelligence, education, and human resources.

The balanced representation of binary classifications for each personality trait ensures an unbiased training ground for machine learning models. Descriptive and visual analyses conducted on the dataset have confirmed its balanced distribution and low inter-trait correlation, fostering confidence in its scientific validity. These features make it an invaluable resource for researchers and practitioners aiming to develop AI systems that can accurately interpret and generate text based on personality traits.

In psychology, this dataset can drive forward the understanding of how personality influences linguistic expression, facilitating more nuanced and personalized therapeutic approaches. For AI and NLP, it offers a structured framework for training models that can emulate human-like text generation conditioned on personality traits, thereby enhancing personalized user interactions in technology-driven applications. Educationally, the dataset supports the creation of personalized learning tools and materials tailored to the diverse personality profiles of students, fostering more inclusive and effective educational environments.

Overall, the introduction of the Big5PersonalityEssays dataset addresses the significant gap in psychometric data availability and paves the way for advancements in both personality research and AI applications. By providing a robust, scientifically validated dataset, this study lays the groundwork for future explorations and innovations at the intersection of these fields. Researchers and developers are encouraged to utilize and build upon this dataset to further their understanding and implementation of personality-aware systems and methodologies.

The dataset is readily accessible for the scientific community via platforms like GitHub, ensuring its availability for widespread use and encouraging collaborative advancements in the field.

Researchers and practitioners using this dataset should ensure appropriate citation of this study to acknowledge its contribution and foster academic integrity.

## 9.Reference List

1. Allport, G. W. (1937). Personality: A psychological interpretation. Holt.
2. Costa, P. T., & McCrae, R. R. (1995). Primary traits of Eysenck's PEN system: Three-factor and five-factor solutions. Journal of Personality and Social Psychology, 69(4), 308-317. https://doi.org/10.1037/0022-3514.69.4.308
3. DeYoung, C. G., Peterson, J. B., & Higgins, D. M. (2002). Higher-order factors of the Big Five predict conformity: Are there neuroses of health? Personality and Individual Differences, 33(4), 533-552. https://doi.org/10.1016/S0191-8869(01)00171-4
4. Digman, J. M. (1997). Higher-order factors of the Big Five. Journal of Personality and Social Psychology, 73(6), 1246-1256. https://doi.org/10.1037/0022-3514.73.6.1246
5. Goldberg, L. R. (1990). An alternative "description of personality": The big-five factor structure. Journal of Personality and Social Psychology, 59(6), 1216–1229. https://doi.org/10.1037/0022-3514.59.6.1216
6. John, O. P., & Srivastava, S. (1999). The Big Five trait taxonomy: History, measurement, and theoretical perspectives. In L. A. Pervin & O. P. John (Eds.), Handbook of personality: Theory and research (Vol. 2, pp. 102-138). Guilford Press.
7. McCrae, R. R., & John, O. P. (1992). An introduction to the five-factor model and its applications. Journal of Personality, 60(2), 175-215. https://doi.org/10.1111/j.1467-6494.1992.tb00970.x
8. OpenAI. (2023). GPT-4 technical report. https://openai.com/research/gpt-4
9. Quinn, P. D. & Duckworth, A. L. (2007). Happiness and the Big Five personality traits: Implications for developing a mindful workforce. Journal of Positive Psychology, 3(3), 114-129. https://doi.org/10.1080/17439760701664654